\documentclass[11pt]{lncs}
\usepackage{mathtools}
\usepackage{amssymb}
\usepackage{graphicx}
\usepackage{amsmath}
\usepackage{comment}
\usepackage{appendix}
\usepackage{subfigure}
\usepackage{float}
\title{Periodicity and perfect state transfer in quantum walks on variants of cycles}
\author{K. Barr, T. Proctor, D. Allen, V. Kendon}
\institute{School of Physics and Astronomy, E C Stoner Building, University of Leeds, Leeds, LS2 9JT, UK}

\begin{document}
\maketitle
\begin{abstract}
We systematically investigated perfect state transfer between antipodal nodes of discrete time quantum walks on variants of the cycles $C_4$, $C_6$ and $C_8$ for three choices of coin operator. Perfect state transfer was found, in general, to be very rare, only being preserved for a very small number of ways of modifying the cycles. We observed that some of our useful modifications of $C_4$ could be generalised to an arbitrary number of nodes, and present three families of graphs which admit quantum walks with interesting dynamics either in the continuous time walk, or in the discrete time walk for appropriate selections of coin and initial conditions. These dynamics are either periodicity, perfect state transfer, or very high fidelity state transfer. These families are modifications of families known not to exhibit periodicity or perfect state transfer in general. The robustness of the dynamics is tested by varying the initial state, interpolating between structures and by adding decoherence. 
\end{abstract}

\section{Introduction}

The development of quantum walks was inspired by classical random
walks. Quantum walks were first introduced by Aharonov \textit{et al}
\cite{Aharonov93} and the concept was developed for specific
applications by Meyer \cite{Meyer961,Meyer962} and Watrous \cite{Watrous01}. The walks were then
examined in their own right by Ambainis \textit{et al} \cite{Ambainis01} on the
line and Aharonov \textit{et al} \cite{AAKV} on a general graph.  Since their introduction, both discrete and
continuous time quantum walks have provided much fruitful research,
see \cite{Kempe03,Kendon07} for overviews. In this paper we
consider two particular aspects of the transport properties of quantum
walks: periodicity and perfect state transfer. The problem of quantum
state transfer was first posed by Bose \cite{Bose03} who considered
using spin chains to transmit quantum states in quantum
computers. Since then, quantum, and in particular perfect state
transfer has been shown to be of interest for other reasons. It is a
necessary ingredient in the quantum walk implementation of the
universal quantum gate set, developed initially by Childs
\cite{Childs09} for the continuous time walk, then by Lovett \textit{et al}
\cite{Lovett10} for the discrete time walk. Quantum walks with good transport properties when undergoing decoherence may provide a model for exciton transport in photosynthetic complexes, as suggested by Mohseni \textit{et al} \cite{Mohseni08}. For modelling charge and
energy transfer, the important factor is total transmitted
amplitude, the phase does not also have to be transmitted.

Most efforts to classify when perfect state transfer can occur have used the continuous time quantum walk model, as the addition of the coin states to the discrete time model makes for more difficult analytical solutions. The purpose of the work presented here is to investigate how strongly the presence of perfect state transfer depends on the precise graph structure that the walk takes place over, and look for particular structural modifications of graphs which do not affect their perfect state transfer properties. Both discrete and continuous time walks were investigated. The focus on small graphs is intended to limit the physical resources required to instantiate the walk, and the families of graphs presented in the main results section are generalisations of some of the small graphs investigated.  

Whether we are looking for perfect state transfer or high amplitude
transfer, the basis of the problem is the same. We wish to find graphs
with pairs of points between which the desired transport takes
place. The graphs considered in this paper are all unweighted and we focus on those built
from some of the simplest families of graphs: complete graphs $K_n$,
cycles $C_n$, and paths $P_n$. As a starting point we focused on variations of cycles known to exhibit perfect state transfer in the discrete time walk. Unweighted graphs are abstractions of
unmodulated spin systems, with the edges representing coupling between
spins. Periodicity is closely related to perfect state transfer. A
graph that has perfect state transfer will also be periodic \cite{Godisil} for any continuous time walk, as well as for discrete time walks over symmetric graphs. 

The paper proceeds as follows: we introduce the relevant background theory relating to quantum walks and graphs in Section \ref{sec:back}; then prior results relating to the graphs of interest in this paper are briefly outlined in Section \ref{sec:prior}; finally in Sections \ref{sec:results1} and \ref{sec:results} we present our results and discuss their robustness. 

\section{Background}
\label{sec:back}
\subsection{Basic Graph Theory}
\label{sec:graphs}
\begin{center}
\begin{figure}[b]
\centering \includegraphics[scale = 0.6]{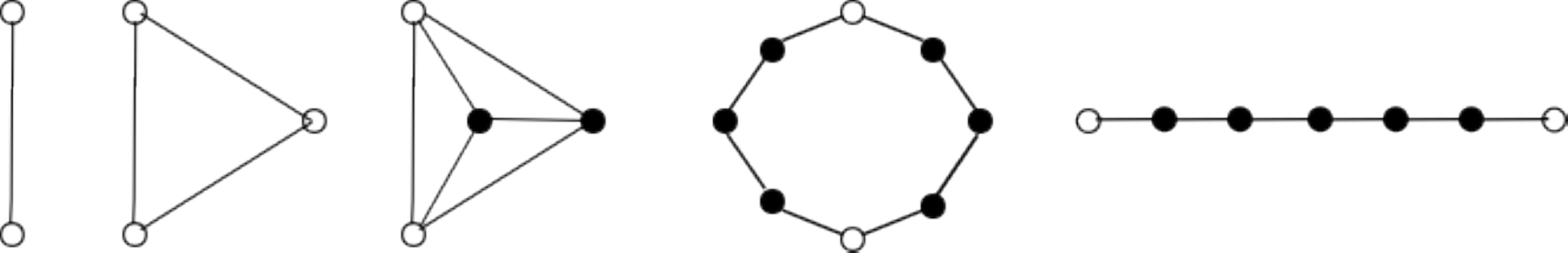}
\caption{Examples of graphs, from left to right: the complete graph
  with 2 vertices, $K_2 = P_2$, the path with 2 vertices; the complete
  graph with 3 vertices $K_3 = C_3$ the cycle with 3 vertices; the
  complete graph $K_4$; the cycle $C_9$ and the path $P_7$. The open
circles indicate potential end points for examining transport properties.} \label{fig:basic}
\end{figure}
\end{center}
\vspace{-8mm}
A graph $\mathcal{G} = \{E, V\}$ is a set of vertices $V$ and a set of edges $E$ of the form $(i,j)$ where $i, j \in V$. The adjacency matrix $A_{\mathcal{G}}$ of $\mathcal{G}$ has ones in the entries $(i, j)$ if vertex $i$ is connected to vertex $j$ and zeroes elsewhere. A graph is \textit{complete} if every vertex is joined by an edge to every other vertex. Some common graphs, which will be used in this paper are $K_n$, the complete graph with $n$ vertices; $P_n$ the path of $n$ vertices and $C_n$, the cycle of $n$ vertices, which differs from the path by only one edge.  Examples of these graphs are shown in Figure \ref{fig:basic}. The \textit{degree} of a vertex is the number of vertices it is joined to. 

The \textit{complement} of a graph $\mathcal{G}$, denoted by $\mathcal{\overline{G}}$, shares the same vertices as $\mathcal{G}$ but its set of edges is $\overline{E} = \{ (i,j) \ | \ (i,j) \notin E \}$. The \textit{join} of graphs $\mathcal{G}$ and $\mathcal{H}$ is denoted $\mathcal{G} + \mathcal{H}$ and its adjacency matrix is given by:
\begin{center}
\begin{equation}
\label{eqn:adj}
A_{\mathcal{G} + \mathcal{H}} = \left( \begin{array}{cc} A_{\mathcal{G}} & J \\ J & A_{\mathcal{H}} \end{array} \right),
\end{equation}
\end{center}
\vspace{5mm}
where $J$ is the all ones matrix of the relevant dimensions. The join therefore links all nodes of $\mathcal{G}$ to all nodes of $\mathcal{H}$. The notion of the adjacency matrix can be generalised to allow it to have entries not equal to either 0 or 1, which varies the strength with which vertices are joined. A graph with only integer values in its adjacency matrix is called an \textit{integral graph}. Classically, integral graphs with adjacency matrices whose entries are greater than one represent multiple edges between nodes. If the diagonals of $A_\mathcal{G}$ are integers, then $\mathcal{G}$ is interpreted as having self loops. Self loops add one to the degree of a vertex.

\subsection{Quantum Walks}
A \textit{discrete time quantum walk} over a graph structure $\mathcal{G} = \{E, V \}$ is controlled by a quantum `coin.' The coin is a unitary operator applied at each vertex $v \in V$, and its dimension is $d \times d$ where $d$ is the degree of the vertex. The state of the quantum walk must describe both the position of the walker, and the configuration of coin states. A general state is written
\vspace{-3mm} \begin{center}
\begin{equation}
\label{discrete}
\psi (T) = \sum_{v,c} \alpha _{v,c} (T) |v, c \rangle
\end{equation}
\end{center}
where $\alpha _{v,c} \in \mathbb{C}$ and $| v, c \rangle$ denotes a
basis state on vertex $v$ with coin state $c$.

After the coin toss, a shift operation is applied. This is simply a permutation between the relevant coin states at different vertices, and is hence  a unitary transformation. The shift operator acts like $\mathcal{S}|v, c \rangle = |w, d \rangle $, so moves amplitude from the $c^{th}$ coin state of $v$ to the $d^{th}$ coin state of $w$ ~\cite{Kendon06}. The probability of the walker being measured at position $v$ after $T$ steps is the summation over coin states at $v$, $ p(v,t) = \sum_{i} |v, c_i |^2$. 

Coin operations which will be particularly relevant to this paper are the Grover operator: 
\begin{center}
\begin{equation}
\label{eqn:grover}
G_{d}= \begin{pmatrix}
 \frac{2-d}{d} & \frac{2}{d} & \cdots & \frac{2}{d} \\
& & &\\
  \frac{2}{d} & \frac{2-d}{d} & \cdots & \frac{2}{d} \\
& & & \\
  \vdots  & \vdots  & \ddots & \vdots  \\
& & & \\
   \frac{2}{d}  & \frac{2}{d} & \cdots & \frac{2-d}{d}
 \end{pmatrix}
\end{equation}
\end{center}
\vspace{5mm}
and the unitary Discrete Fourier Transform (DFT), specified by:
\begin{center}
\begin{equation}
\label{eqn:DFT}
DFT =  \frac{1}{ \sqrt{d}}\begin{pmatrix}
\omega^{0 \times 0}_d & \omega^{0 \times 1}_d & \ldots & \omega^{0 \times  (d-1)}_d \\
\omega^{1 \times  0}_d & \omega^{1 \times  1}_d & \ldots & \omega^{1 \times (d-1)}_d \\
 \vdots  & \vdots  & \ddots & \vdots  \\
\omega^{(d-1) \times  0}_d & \omega^{(d-1) \times  1}_d & \ldots & \omega^{(d-1) \times (d-1)}_d \\
\end{pmatrix}
\end{equation}
\end{center}
\vspace{5mm}
with $\omega_d = e^{\frac{- 2 \pi i}{d}} $. The $d = 2$ version of the DFT is more commonly known as the Hadamard, $H$. If the columns of $H$ are swapped around we have $H_2$:

\hspace{30mm} \begin{equation}
\label{eqn:h2}
H =\frac{1}{\sqrt{2}} \begin{pmatrix} 1 & 1 \\ 1 & -1 \end{pmatrix}  
\hspace{25mm}  H_2 = \frac{1}{\sqrt{2}} \begin{pmatrix} 1 & 1 \\
-1 & 1
\end{pmatrix}
\end{equation}
Various attributes of discrete time quantum walks are of potential interest, such as the limiting distribution, \cite{Tregenna03,AAKV}, and hitting time, \cite{Krovi,Krovi06,Kempe05}. The aspect of interest for this paper is whether it admits perfect state transfer, defined in Section \ref{sec:pst} below.

A \textit{continuous time quantum walk} evolves over graph
$\mathcal{G} = \{ E, V \}$ according to the Schr\"{o}dinger equation. The adjacency matrix $A_\mathcal{G}$ gives rise to the
Hamiltonian, which is taken to be $\mathcal{H} = \gamma A_{\mathcal{G}}$ where
$\gamma$ is the hopping rate. As there is no coin space, the walk evolves entirely in
the position basis. These walks were introduced by Farhi and Gutmann
\cite{FarhiGutt} who showed they exhibited exponentially faster transport over 
glued binary trees than the classical versions. This walk was analysed
further by Childs \textit{et al} in \cite{Childs02}. 

In units where $\gamma = \hbar = 1$, the state of a continuous time walk at time $t$ is given by
\begin{center}
\begin{equation}
\label{eqn:ctimepsi}
| \phi (t) \rangle = e^{-i A_{\mathcal{G}} t} | \phi (0) \rangle.
\end{equation}
\end{center}
\vspace{5mm}
The amplitude at vertex $v$ at time $t$ is $\langle v | \phi (t)
\rangle$. 

Unlike in the discrete time case, where multiple coin states can accommodate
multiple edges between the same two vertices, there can be at most one edge
between vertices of $\mathcal{G}$ in the continuous time case. Varying
the value of the entry $A_\mathcal{G} (i,j)$ gives rise to a weighted
edge between vertex $i$ and vertex $j$, which is equivalent to repeating an edge when the value is made greater than one. 

\subsection{Perfect State Transfer Conditions}
 \label{sec:pst}
 
For the continuous time walk, there is \textit{perfect state transfer} between vertices $v$ and $w$ at time $t$ if the following condition is met:
\begin{center}
\begin{equation}
\label{eqn:cpst}
\langle w | e^{-i A_\mathcal{G} t} | v \rangle = 1
\end{equation}
\end{center}
\vspace{5mm}
where $|v \rangle$ and $|w \rangle$ are unit vectors at vertices $v$
and $w$ respectively and $t > 0$. If the perfect state transfer
condition holds for some $v,w \in V$ then we say the graph $\mathcal{G}$ has perfect state transfer. A special case of perfect state transfer occurs when $v = w$, in which case the graph $\mathcal{G}$ is \textit{periodic}.

In the discrete time quantum walk, \textit{perfect state transfer} occurs between vertices $v$ and $w$ after $T$ steps if
\begin{center}
\begin{equation}
\label{eqn:dpst}
\sum_{c,d} \langle w, d | (\mathcal{SC})^T | v, c \rangle = 1.
\end{equation} 
 \end{center}
 
It is clear from this definition that we are not concerned about
whether the coin states are in the same configuration at vertex $w$ as
they were at vertex $v$, which enables perfect state transfer between
vertices of different degrees. For the definition of periodicity, the
stricter requirement of returning to the same configuration of coin
states is made:

\begin{center}
\begin{equation}
\label{eqn:dpst}
\sum_{c} \langle v, c | (\mathcal{SC})^T | v, c \rangle = 1.
\end{equation} 
\end{center}

The perfect state transfer conditions may be stricter than necessary for some
applications, in which case one can look for \textit{high amplitude transfer}
instead. A lower bound $\lambda$ for the transmitted amplitude must
be selected. For the numerical work in this paper we take $\lambda = 0.9$, and we require that 

\begin{center}
\begin{equation}
\begin{aligned}
\sum_{c,d} \langle w, d | (\mathcal{SC})^T | v, c \rangle \geq
\lambda & \hspace{10mm} \langle w | e^{-i A_\mathcal{G} t} | v \rangle \geq \lambda 
\end{aligned}
\end{equation} 
\end{center}
in the discrete and continuous time cases respectively. 

\subsection{Decoherence in Quantum Walks}

Decoherence is modelled in the standard way, using projection
operators. A detailed account of how to model uncorrelated environmental interactions in various
types of quantum walks can be found in Kendon \cite{Kendon07}. For the discrete time quantum walks, there are two potential types of decoherence: in the coin basis and in the position basis. In both cases the density operator notation must be used, so the state of the quantum walker is written
\begin{center}
\begin{equation}
\label{eqn:density1}
\rho = \sum_{v, c} \sum_{v',c'} = \rho_{v,c, v', c' } |v, c \rangle \langle v', c' |
\end{equation}
\end{center}
 
Decoherence in either basis can be modelled using a map
\begin{center}
\begin{equation}
\label{eqn:ddec}
\mathcal{U}: \rightarrow \sum_{j \in \Theta} \mathbb{P}_j \mathcal{SC} \rho \mathcal{C}^{\dagger} \mathcal{S}^{\dagger} \mathbb{P}^{\dagger}_j 
\end{equation}
\end{center}
\vspace{5mm}
where the projection operators $\mathbb{P}_j$ can project either in
the coin basis, the position basis, or both. The set of projection
operators $\Theta$ has cardinality one if the evolution is unitary, as
$\mathbb{P}_j $ must be the identity in this case. The evolution of the quantum walk, with probability of a decoherence event occurring during a given timestep $p$ is then given by 
\begin{center}
\begin{equation}
\label{eqn:dtimeev}
\rho (t+1) = (1-p)\mathcal{SC} \rho \mathcal{C}^{\dagger} \mathcal{S}^{\dagger} + p \sum_{j} \mathbb{P}_j \mathcal{SC} \rho \mathcal{C}^{\dagger} \mathcal{S}^{\dagger} \mathbb{P}^{\dagger}_j.
\end{equation}
\end{center}
 
In the continuous time case, the purely quantum time evolution in density matrix notation is given by 

\begin{center}
\begin{equation}
\label{eqn:contdens}
\frac{d \rho(t)}{dt} = -i [A_{\mathcal{G}}, \rho]
\end{equation}
\end{center}
 \vspace{5mm}
 and when there is Markovian noise present the density matrix evolves according to 
 
 \begin{center}
\begin{equation}
\label{eqn:contdec}
\frac{d \rho (t)}{d t} = -i [ A_{\mathcal{G}}, \rho (t)] - p \rho (t) +  p \sum_j \mathbb{P}_j \rho (t) \mathbb{P}^{\dagger}_j
\end{equation}
\end{center}
 
The effect of decoherence in both walks is that the off-diagonals of
the density matrix, the coherences, decay. In the long time limit, or with decoherence rate $p=1$, the classical walk is recovered. This property allows us to consider the quantum walk to be the quantum analogue to the classical random walk. 

\section{Prior Results}
\label{sec:prior}
The graphs discussed in Section \ref{sec:results} below are composed
from the graphs $\overline{K_n}$, the complement of the complete graph with $n$ nodes, specifically $\overline{K_2}$; $P_n$, the path with $n$ nodes and $C_n$, the cycle with $n$ nodes. In this section, previous results concerning these graphs are discussed. It is clear
that as there are no edges joining the vertices of $\overline{K_n}$,
quantum walks on this structure do not evolve in the position
basis. The complete graph $K_n$ has been shown not to exhibit perfect
state transfer, but this can be achieved by the removal of an edge \cite{Bose09}.

\subsection{Discrete Time Quantum Walk}
\label{sec:dwpres}
The quantum walk on the path $P_n$ with $n= \infty$, or $n>2T$ if the amplitude is initially situated at the middle vertex, is simply the quantum walk on the line \cite{Ambainis01}. The dynamics of the walk on the line are well known. The effect of varying the initial coin state, and coin operator, is investigated analytically by Bach \textit{et al} in \cite{Bach} and discussed by Tregenna \textit{et al} in \cite{Tregenna03}. It is found that for a given coin, the whole range of dynamics available to the quantum walk on the line can be observed by varying the initial state. The evolution on the infinite line is solved exactly by Ambainis \textit{et al} \cite{Ambainis01}. The most notable fact about the walk on the line is that the standard deviation of the position of the walker varies proportionally to $T$, as opposed to with $\sqrt{T}$ in the classical case.

Discrete time quantum walks along finite paths $P_n$ do not in general admit perfect state transfer for non trivial coin operators, i.e., those not equal to the Pauli $X$ operator, which is also the $2$ dimensional Grover operator. Inspired by the results from \cite{Christandl04} indicated in Section \ref{sec:prct} below, for the continuous time case, we added self loops to the ends of paths to see if this improved transport. Though adding self loops to the ends of $P_4$ created a walk with perfect state transfer, and hence due to symmetry, periodicity, adding self loops in general did not improve transport across the paths. 

As very few cycles admit perfect state transfer for either the continuous, \cite{Saxena07,Basic09}, or discrete time walk, \cite{Travaglione02}, the dynamics on $C_n$ is more often analysed in terms of mixing times, which are not of interest in this paper. 

Periodicity in cycles was first noted for $C_4$ \cite{Travaglione02}. In cycles with even $n \leq 10$ and a suitable
coin, periodicity is observed and perfect state transfer occurs
halfway through the period. Varying $n$ can have a dramatic effect on
the dynamics of the walk, doubling it to go from $n=8$ to $n=16$ turns
periodic evolution into highly irregular evolution. 

The effects of decoherence in both the walk on the line and walks on
various cycles were examined by Kendon and Tregenna in \cite{Kendon03},
where they conclude that decoherence at a suitable rate enhances some
qualities of the walk. Decoherence is shown to give rise to a uniform
distribution if the correct rate for the number of timesteps is used,
as discussed in further detail by Maloyer and Kendon,\cite{Maloyer07}. The distribution generates a central 'cusp' if the decoherence is in the coin basis \cite{Dur}. The sensitivity of the walk to decoherence at rate $p$ grows linearly with the number of timesteps $T$ that the walk undergoes. 
\subsection{Continuous Time Quantum Walk}
\label{sec:prct}

The investigation of perfect state transfer along paths was initiated
by Bose \cite{Bose03} where a string of coupled qubits was
considered. Later results, by Christandl \textit{et al} \cite{Christandl04}, show that there is no perfect state transfer between antipodal points on paths of length $\geq 4$ with unweighted edges. 

Cycles are a type of graph known as an integral circulant. Integral
circulants were first examined by So in \cite{So05}, and results
concerning their perfect state transfer properties were developed by
Saxena \cite{Saxena07} and built upon by Ba\v{s}i\'{c}
\cite{Basic09}. Integral circulants with odd numbers of nodes cannot
have perfect state transfer \cite{Saxena07}. The only $n$-even cycle with perfect state transfer is $C_4$, in contrast to the discrete time case where it is exhibited up to $C_{10}$. 

Kendon and Tamon  \cite{PSTrev} review many results, both analytic and numeric,
concerning perfect state transfer in the continuous time walk, as well as
some for the discrete time case. Kay \cite{Kay10}
reviews the necessary and sufficient conditions for systems with
nearest neighbour interactions to exhibit perfect state transfer and
how these can be used to design systems which implement other
protocols.

\section{Discrete time quantum walks over small structures}
\label{sec:results1}

We did numerical simulations of discrete time quantum walks over a variety of structures based on graphs known to exhibit perfect state transfer under some circumstances, in order to see if related structures displayed similar properties. All of the walks were run for 100 steps using 3 types of coin operator: 
\begin{center}
\begin{itemize}
\item{$\mathcal{O}_1 =$ DFT at every node}
\item{$\mathcal{O}_2 =$ Grover at every node}
\item{$\mathcal{O}_3 =$ $H$ at nodes of degree two and Grover at other nodes}
\end{itemize} 
\end{center}
$\mathcal{O}_2$ and $\mathcal{O}_3$ will be identical in graphs with no vertices of degree two. To assess the sensitivity of perfect state transfer to the initial state, walks using 1500 initial states with different configurations of coin states at a particular node, uniformly distributed according to the Haar measure \cite{Nemoto00}, were run for 100 steps. This number of steps is sufficient to determine whether behaviour such as periodicity is exhibited for graphs of this size. The number of initial conditions was selected after preliminary investigations revealed that increasing the number of initial conditions beyond 1500 did not affect the results. 

\subsection{Diamond Chains}
\label{sec:diamond}
\begin{figure}  \subfigure a) {\includegraphics[scale = 0.78]{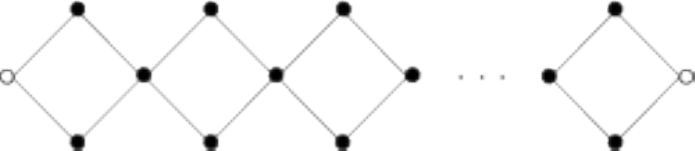}} \subfigure b) {\includegraphics[scale = 0.78]{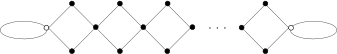}} \\
 \subfigure  c) {\includegraphics[scale = 0.78]{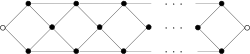}} \subfigure d) {\includegraphics[scale = 0.78]{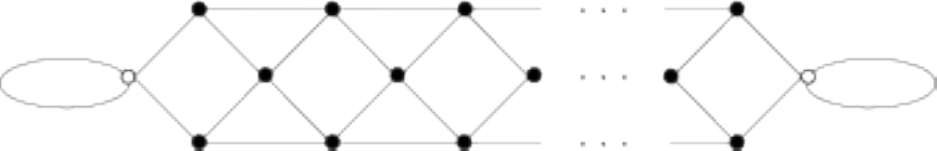}} \caption{The basic diamond chain, (a), and the variants investigated. Vertices with self loops have degree 3.} \label{fig:diamonds}\end{figure}

Chains of $n$ diamonds where all amplitude is initially equally distributed between the coin states at an end vertex are known to exhibit perfect state transfer after $2n$ steps when the Grover coin operator is used \cite{PSTrev}. This is because on structures of even degree, if the amplitude is in initially equally distributed between half of the coin states, then it will be transferred into the remaining coin states. This observation enables many other structures exhibiting perfect state transfer to be specified. For example, the graph formed by combining chains of diamonds to create a rectangular structure will exhibit perfect state transfer if all amplitude is initially equally distributed between the coin states at the edge of the structure. Three variants of the diamond chain, shown in Figure \ref{fig:diamonds}, were investigated.

The perfect state transfer in the diamond chain was found to be highly dependent on both the structure and the choice of coin operator. For operators $\mathcal{O}_1$ and $\mathcal{O}_3$, no perfect state transfer occurs. The only variant of the chain tested found to have high amplitude transfer, with cutoff 0.9, was that depicted in Figure \ref{fig:diamonds} (b), with self loops at either end of the chain. This transfer occurred for only 35 out of 1500 initial states tested, so even the addition of a single edge to the end nodes along the chain destroys the perfect state transfer. This is because the additional coin state at the initial node makes achieving the required equal superposition between the other two coin states impossible. If an additional self loop is added, then the initial node has even degree so perfect state transfer would be recovered. The other variants of the diamond chain did not exhibit perfect state transfer or periodicity for any selection of operator tested, and the maximum probability observed at the end of the chain decreased roughly as $n$ increased. The results are summarised in Appendix \ref{sec:appa}.

\subsection{Variants of Cycles}
\label{sec:cycles}
Very few simple structures are known to exhibit perfect state transfer in the discrete time walk \cite{PSTrev} so more complex structures were investigated. As noted in Section \ref{sec:dwpres}, even cycles with $n \leq 10 $ have periodicity. In particular, cycles $C_4$ and $C_8$ are periodic when the Hadamard operator is used for the coin, and perfect state transfer occurs half way through the period. The cycle $C_6$ is periodic when a biased coin operator is used \cite{Tregenna03}. In order to test the sensitivity of this perfect state transfer to the structure we first investigated how a small set of specific modifications effects the perfect state transfer, none of these admitted perfect state transfer, the results are briefly outlined in Appendix B. 

\begin{figure}  \subfigure a) {\includegraphics[scale = 0.12]{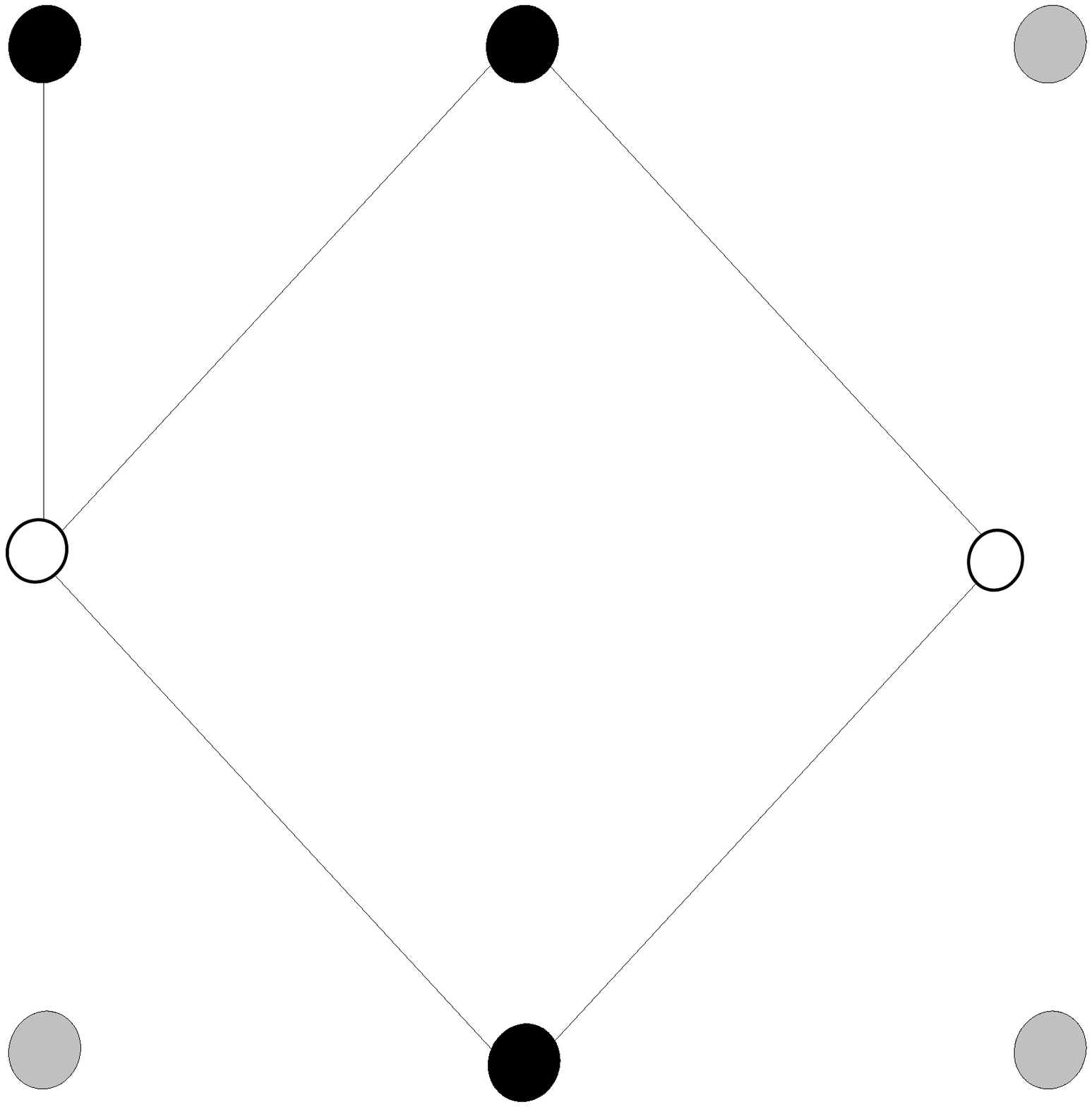}} \subfigure b) {\includegraphics[scale = 0.12]{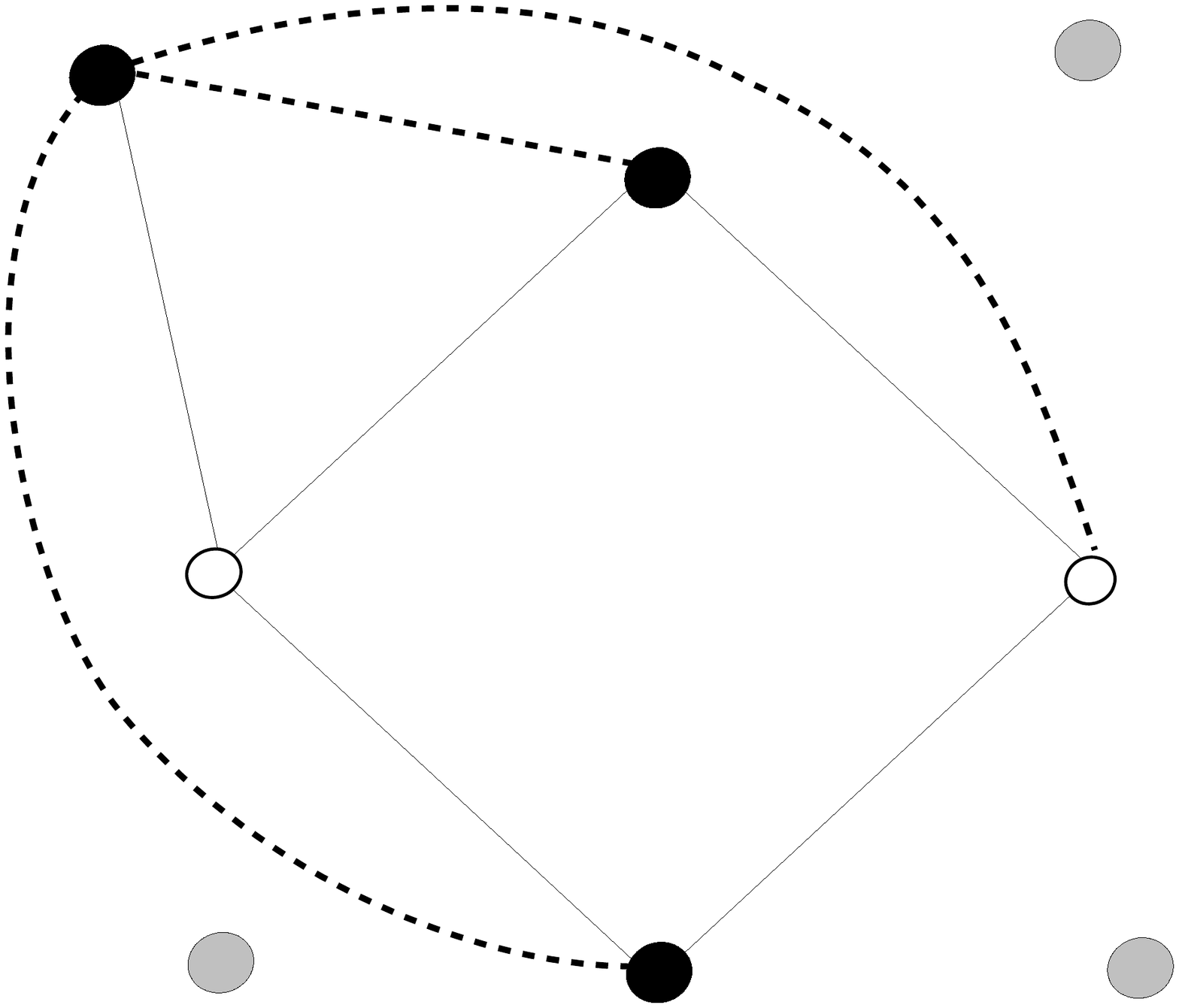}} \subfigure  c) {\includegraphics[scale = 0.12]{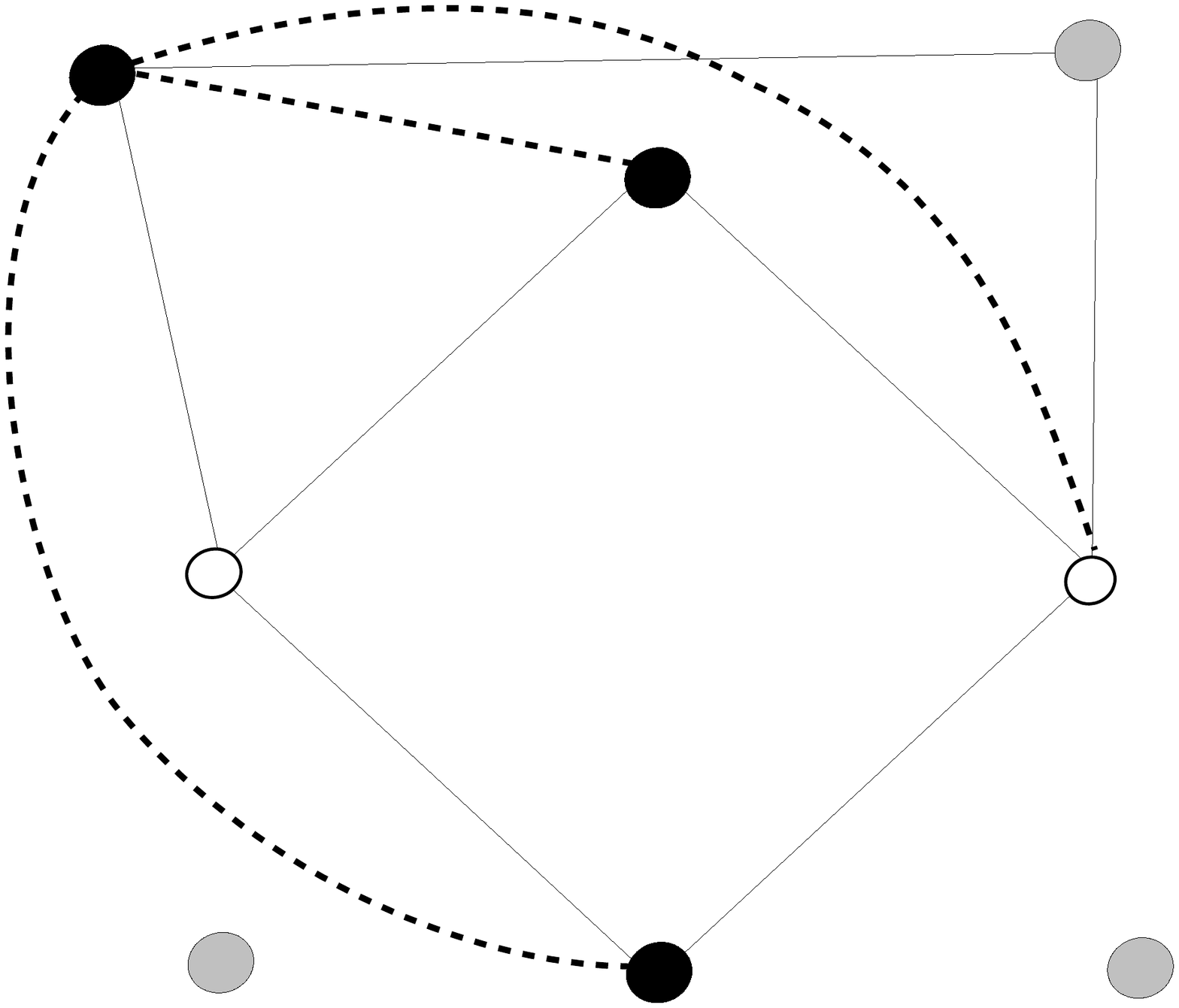}} \caption{Generation of the variants of cycles, shown in the case of $C_4$: a) attach individual node to an existing node b) attach new node to every existing node c) combine each of these graphs, and add links between new nodes. Dotted lines indicate other connections added for a single node, the rest are omitted for clarity} \label{fig:systematic}\end{figure}

Following this preliminary investigation we systematically simulated walks over structures based on $C_4, C_6$ and $C_8$. We modified the structures by adding up to four new nodes. This modification worked in the following way: First, a single node was added, in turn, to each existing node of the cycle. Technically this creates the same graph each time, but as we always started with amplitude at the same node, the dynamics of the walk will be different depending on the node of the cycle that the new node has been added to. We then generated further edges between that new node, and the nodes of the existing cycle. After structures including each new node added to the cycle in every possible way, see Figure \ref{fig:systematic} (b), it was necessary to combine these structures so that the new nodes were joined not only to the cycles, but to each other. This was performed by taking the sets of adjacency matrices which join each new node to the existing cycle (which are technically for identical graphs, but when combined will lead to new structures) and combining each of them in turn, then joining the new nodes to each other, see Figure \ref{fig:systematic} (c). The authors do not claim that this is the most efficient way to generate these structures. As many duplicates of adjacency matrices, which were removed before simulations were run, and permutations of the same graph are created, however we were more concerned by implementing a systematic study rather than a computationally efficient one.  

Operator $\mathcal{O}_2$ resulted in the most walks with high amplitude transfer, this is due to the fact that it simply performs a swap at nodes of degree 2 and the graphs are symmetrical. Some variants of cycles found to have perfect state transfer between antipodal nodes are examples of the three families of graphs discussed in Section \ref{sec:results}. As we were particularly interested in perfect state transfer between antipodal nodes, we now turn to the results relating specifically to this aspect of the walks in turn. The limitation of our attention to perfect state transfer between antipodal nodes was motivated by further preliminary studies where an arbitrary node was examined instead. Whilst this study was not sufficiently systematic to make any conclusive remarks (using all examined variants of $C_6$ and $C_8$ and examining a single node), it yielded no perfect state transfer. 


\subsubsection{Results}

\begin{figure} \subfigure a) {\includegraphics[scale=0.12]{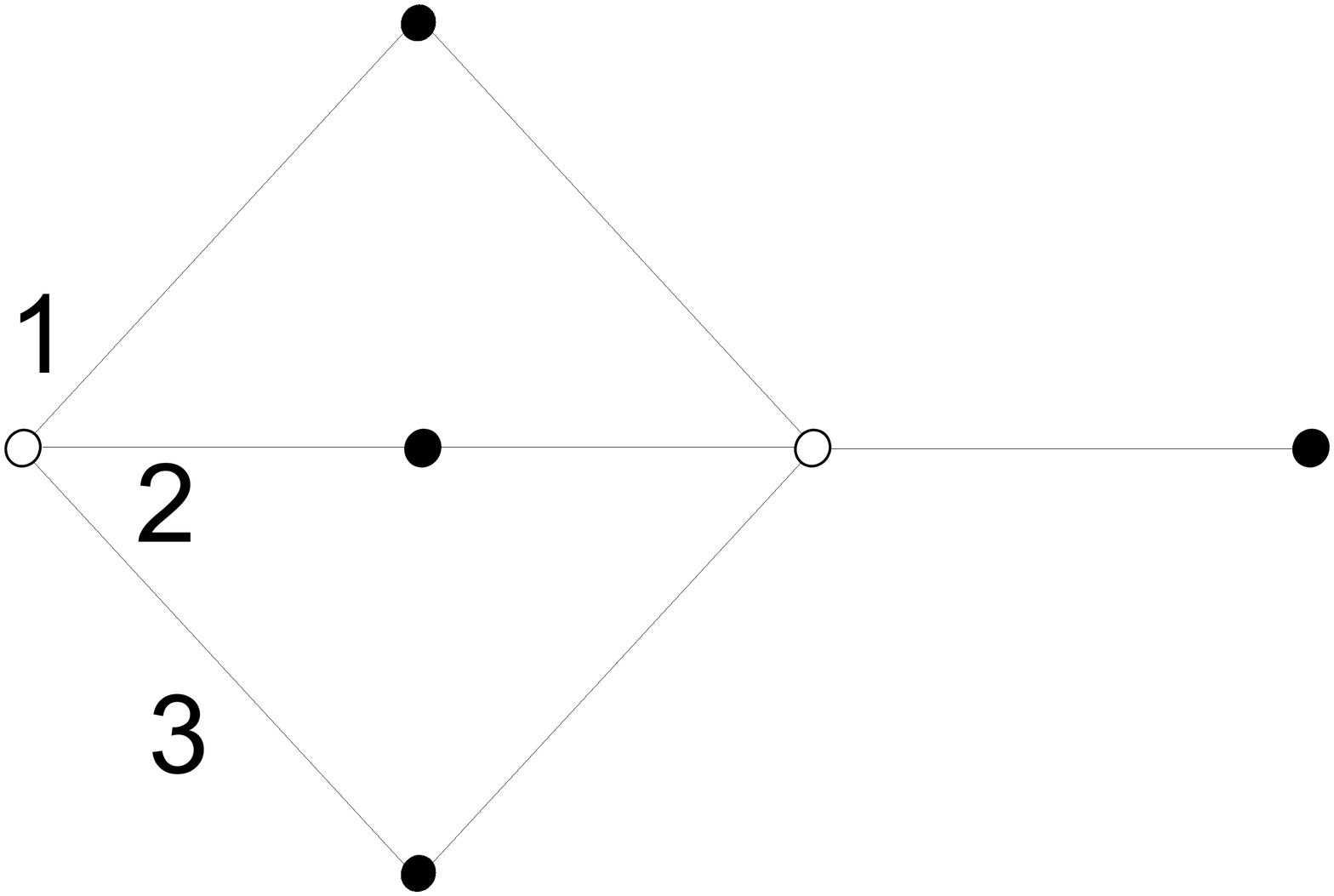}} \subfigure b) {\includegraphics[scale=0.12]{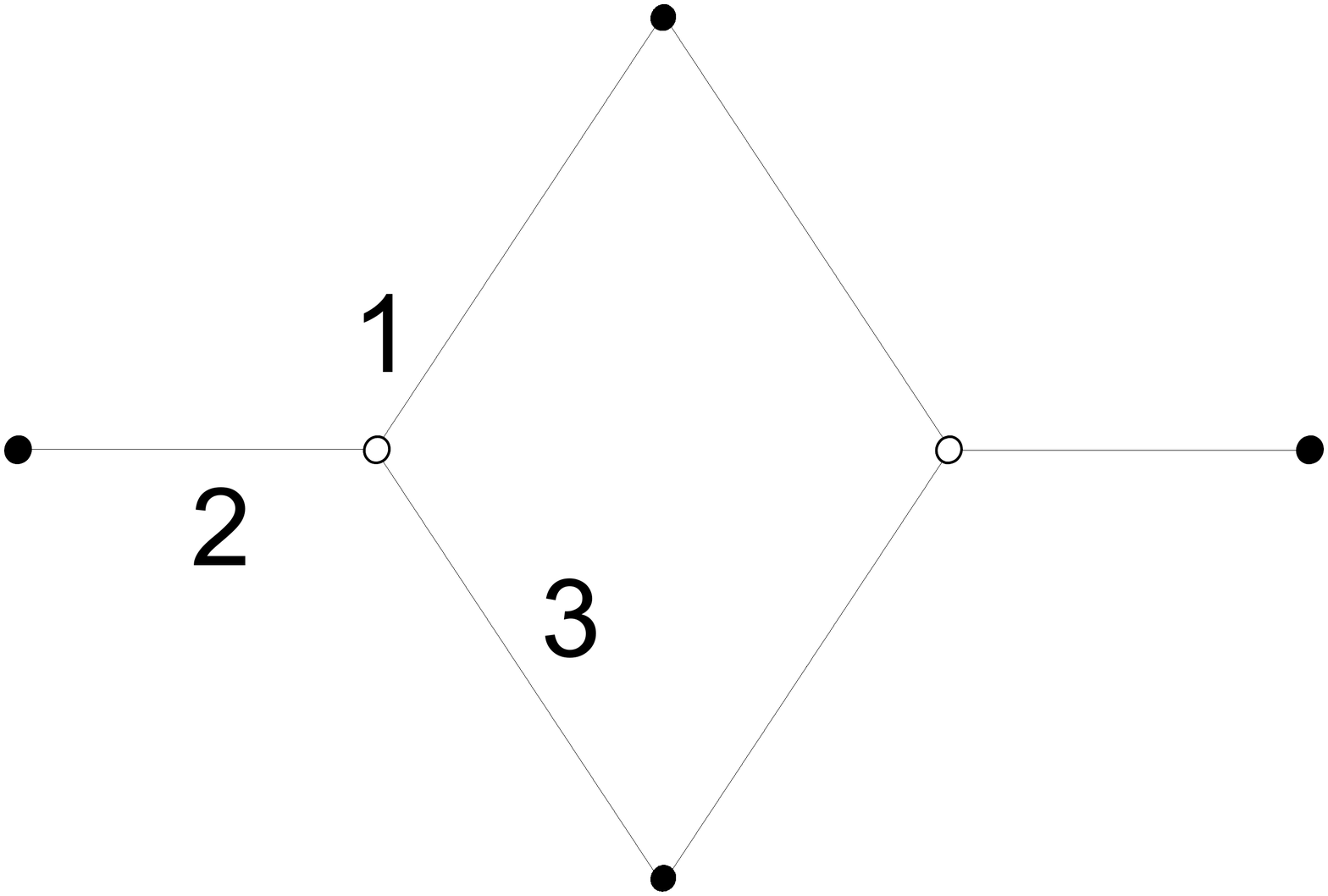}} \subfigure c) {\includegraphics[scale=0.12]{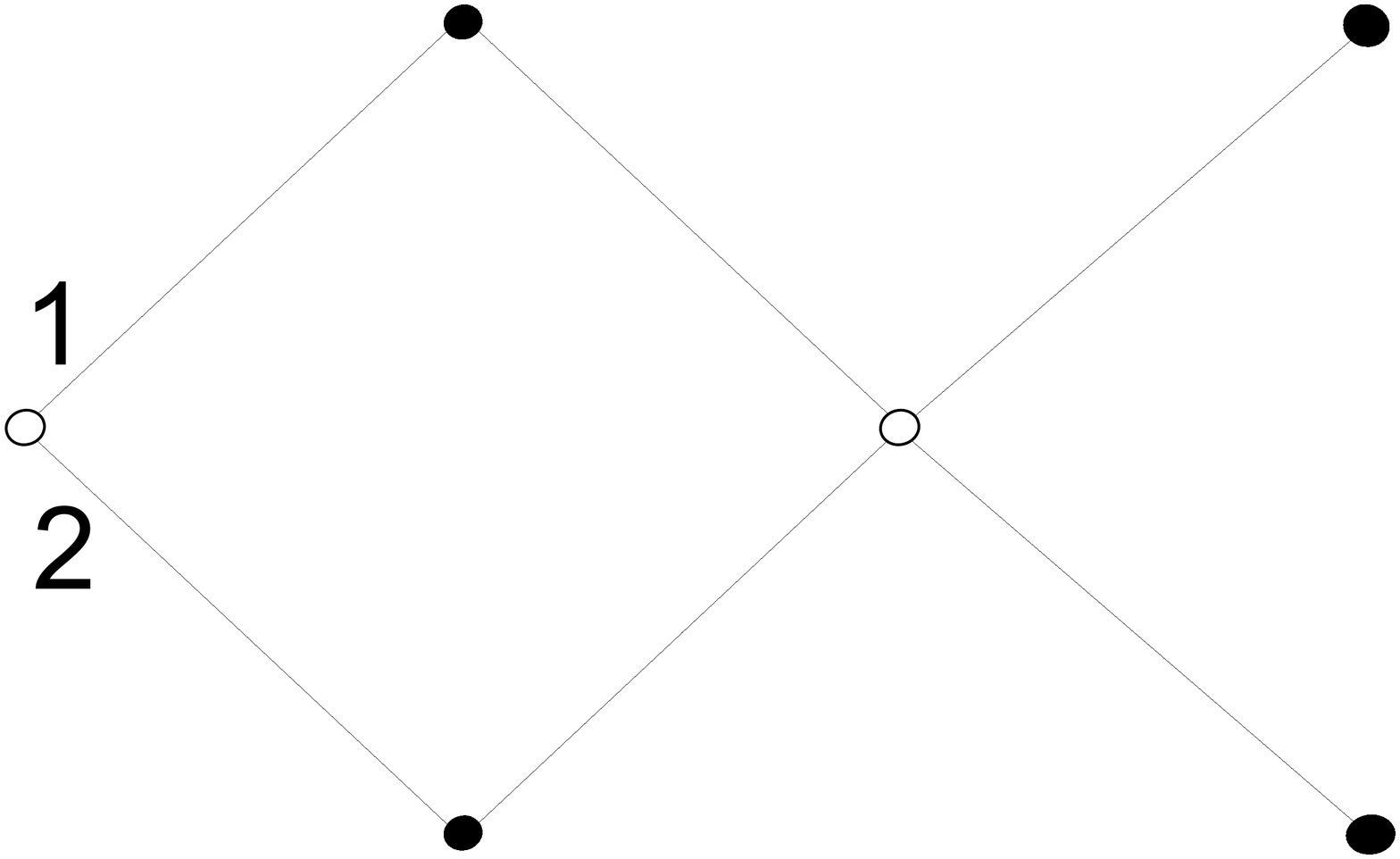}} \caption{Variants of $C_4$ leading to non-trivial perfect state transfer between antipodal nodes. The edge labels correspond to the coin states they represent when the state at that vertex is written as a vector} \label{fig:psts4cycle} \end{figure}

The results were very similar for $C_4$, $C_6$ and $C_8$ so we discuss them together. The only operator we tested leading to perfect state transfer on the variants of cycles was $\mathcal{O}_2$. As it is clear that any variant of an even cycle which adds nodes only to the target node will have perfect state transfer after a number of steps determined by the size of the cycle using this operator, we do not discuss these results. In fact, the only variants of cycles found to exhibit perfect state transfer were those with modifications to the antipodal nodes. As it is already known that on structures with even degree, if amplitude populates half the coin states and is equal in magnitude and phase, then the Grover operator perfectly transmits the amplitude to the other half of the coin states \cite{PSTrev}, walks which reproduced these results are also not discussed. Two sets of graphs based on $C_4$ found to exhibit perfect state transfer can be generalised to families, these are discussed in detail in Section \ref{sec:results} below. After pruning and generalising the results, this leaves us with three new variants of $C_4$ exhibiting perfect state transfer, shown in Figure \ref{fig:psts4cycle}. Two of the variations also lead to perfect state transfer in $C_6$ and $C_8$ for the correct choices of initial conditions. The low number of positive results implies that perfect state transfer is heavily sensitive to the graph structure used for the walk. The further analysis on these variants has been performed analytically, rather than numerically.

Perfect state transfer occurs for the graph shown in Figure \ref{fig:psts4cycle} (a) with initial states:

\begin{center}
\begin{math}
| \psi_{+, -} \rangle = \begin{pmatrix}  0 \\ (-) \frac{1}{\sqrt{2}} \\ (-) \frac{1}{\sqrt{2}} \end{pmatrix}
\end{math}
\end{center}
Where $\psi_{+}$ has a minus sign for the first populated coin state and $\psi_{-}$ for the second. The coin states are represented in the order shown in Figure \ref{fig:psts4cycle}. This perfect state transfer occurs after 50 steps. As the coin states at the target node are not identical to the initial state, this perfect state transfer does not lead to periodicity.

For the graph shown in Figure \ref{fig:psts4cycle} (b) the following initial states lead to perfect state transfer after 20 steps:

\begin{center}
\begin{math}
| \psi_{+,-} \rangle = \begin{pmatrix} (-) + 0.705 \\ (+) -0.709 \\ (+) - 0.005 \end{pmatrix}
\end{math}
\end{center}
Where the unbracketed signs refer to $|\psi_+ \rangle$ and those in brackets occur in $| \psi_- \rangle$ and the coin states are represented in the order indicated on the figure. The decimal expansions for this set of initial conditions do not appear to have any simple algebraic forms. Again, as the coin states at the target node do not replicate the initial conditions, this perfect state transfer does not lead to periodicity. Additionally this variation admits perfect state transfer for any even cycle after the correct number of steps if the initial conditions are selected such that after the coin flip, only the coin states directing amplitude around the cycle are populated. If we call the amplitude in the coin states on the cycle $x$ and $y$, and require that $xx^* + yy^* = 1$ then the initial states leading to this perfect state transfer can be written:

\begin{center}
\begin{math}
| \psi \rangle = \begin{pmatrix} \frac{2y - x}{3} \\ \frac{2x-y}{3} \\ 2\{ \frac{2x-y}{3} + \frac{2y-x}{3} \}  \end{pmatrix}
\end{math}
\end{center}

For example, if we select $x = y = \frac{1}{\sqrt{2}}$ then the initial condition leading to perfect state transfer will be:

\begin{center}
\begin{math}
| \psi \rangle = \begin{pmatrix}
\frac{1}{3} \frac{1}{\sqrt{2}} \\
\frac{1}{3} \frac{1}{\sqrt{2}} \\
\frac{4}{3} \frac{1}{\sqrt{2}} 
\end{pmatrix}
\end{math}
\end{center}

This case is one of an entire class of cases, another example would be joining antipodal nodes by a single edge, where the initial condition is selected in order to guarantee that no amplitude is in the coin state for the new edge after the coin flip. In other words, the initial condition is selected so that the new structure does not affect the evolution. Clearly this will only preserve the initial perfect state transfer achieved by deterministically traversing the cycle, and after this, the additional coin states will affect the evolution. 

The final graph we discuss, in Figure \ref{fig:psts4cycle} (c), exhibits perfect state transfer for all initial conditions. We denote these $a$ and $b$ as there are two coin states at the initial node. There is the expected perfect state transfer after two steps. Then after ten steps the state at the target 
node is:

\begin{center}
\begin{equation}
\label{eqn:tails}
| \psi_{target} \rangle = \begin{pmatrix} \frac{1}{2}b - \frac{1}{2}a \\ \frac{1}{2}a - \frac{1}{2}b \\ \frac{1}{2}a + \frac{1}{2}b \\ \frac{1}{2}a + \frac{1}{2}b \end{pmatrix}
\end{equation}
\end{center}

Unlike the other walks found, this walk is periodic, after a further two steps the initial condition is recovered. As the form of Equation \ref{eqn:tails} does not depend on the length of the cycle, any even cycle with this modification will lead to perfect state transfer, with a corresponding scaling of period and perfect state transfer time. 

\subsubsection{Continuous time walk on these structures} 
Whilst the purpose of this section of the paper is to investigate perfect state transfer in the discrete time walk over small structures, simulations of the corresponding continuous time walks were also run. In the continuous time walk considerations regarding numerical accuracy are more important. To this end, results from simulations obtained from Python were compared to the corresponding simulations in MATLAB. The results of this comparison, where 1.000 in Python became 0.997 in MATLAB, indicate that our methods are not suitable for studying perfect state transfer in the continuous time walk. However, the methods used can help in narrowing down which graphs to look for perfect state transfer in. For example, the families in Section \ref{sec:results} were highlighted by these simulations, and with further analytic work we were able to prove in one case that perfect state transfer did take place. As the cut-points used to test for perfect state transfer in were the same in the discrete and continuous time case, we were able to draw one comparison- namely that very high amplitude transfer is more common in the continuous time walk than the discrete time walk, for instance being admitted by 16 variants of $C_4$. Whilst there were commonalities between these 16 variants, such as the majority having a constant number of nodes along any path joining the initial and target node, without knowing precisely whether they admit perfect state transfer concrete conclusions cannot be drawn.  
 




\section{Three related families of graphs}
\label{sec:results}
\vspace{3mm}

\begin{figure}[ht] 
\centering 
\subfigure {$\overline{K_2} + \overline{K_n}$} {\includegraphics[scale = 0.4]{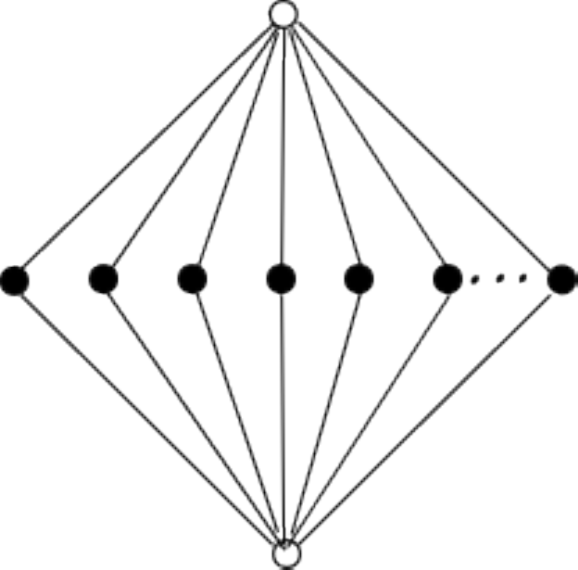}} 
\subfigure {$\overline{K_2} + P_n$} {\includegraphics[scale = 0.4]{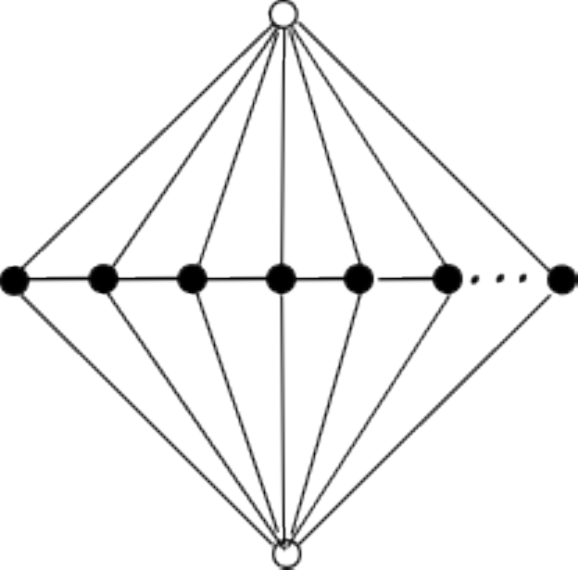}}  
\subfigure {$\overline{K_2} + C_n$} {\includegraphics[scale = 0.4]{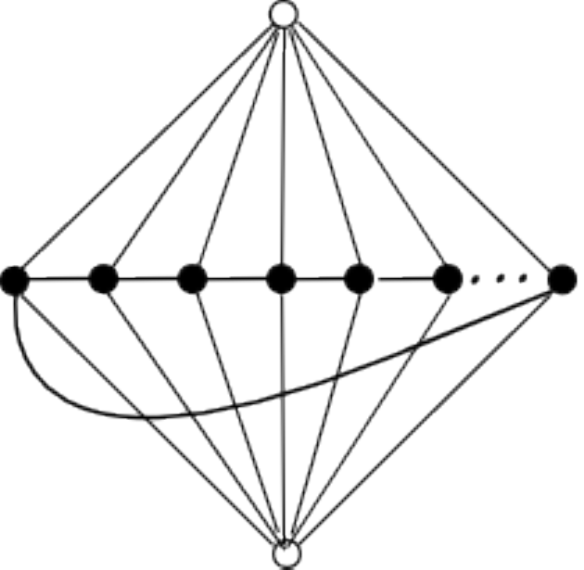}} 
\caption{The three families. The nodes highlighted with open dots indicate the initial and target nodes, due to the symmetry of the graphs, it does not matter which is which.} \label{fig:3fams} \end{figure}


In this section we discuss three families of graphs depicted in Figure \ref{fig:3fams}. These have the same number of nodes, and all include a join with the graph $\overline{K_2}$. The family $\overline{K_2} + P_n$ has no notable transport properties in the discrete time walk, but exhibits perfect state transfer in the continuous time walk for some choices of $n$.

\subsection{Periodicity and perfect state transfer on the graph $\overline{K_2}+\overline{K_n}$}
\label{sec:k2kn}
Due to the fact that the 2 dimensional Grover coin is simply a swap operator, it is obvious that $\overline{K_2}+\overline{K_n}$ exhibits perfect state transfer between the nodes of $\overline{K_2}$ in two steps when the Grover coin is used at all nodes and all amplitude is initially at one vertex of $\overline{K_2}$. For this state transfer, it does not matter how the amplitude is distributed between the coin states at the initial vertex, or whether any of this amplitude has an associated phase. 

Periodicity can be achieved using less trivial operations, as outlined for various cases in Table \ref{tab:table} below. Perfect state transfer to the target vertex is not observed half way through the period of the walk, except in special cases, usually with $n =2$ where the graph is equal to $C_4$. If the DFT coin is used at the nodes of $\overline{K_2} + \overline{K_n}$, but with half of the Hadamards replaced by $H_2$, all the amplitude returns to the initial vertex when $t$ is half the period, with the state being the DFT of the initial state.
\begin{table}[b]
\begin{tabular} {|p{7cm} | p{4cm} | p{1.5cm} |}
\hline
\textit{Coins used at each vertex}  & Initial state used & \textit{Period} \\
\hline
\textbf{All DFT} & Any & 8\\ \hline
\textbf{Any unitary at $\overline{K_2}$, $G_2$ in $\overline{K_n}$} & Any & 4 \\ \hline
\textbf{$G_n$ at $\overline{K_2}$, $H$ at $\overline{K_n}$} & Any & 4 \\ \hline
\textbf{$G_n$ at $\overline{K_2}$, $H$, $H_2$ at half nodes each, $n>2$, even} & See below* & 8 \\
\hline
\end{tabular} 
\caption{The periods of walks with various coins over $\overline{K_2}+\overline{K_n}$ with all amplitude initially in one of the vertices of $\overline{K_2}$. *Equal superposition over all coin states and all amplitude initially in one coin state.}
\label{tab:table}
\end{table}

\subsection{Periodicity and perfect state transfer in $\overline{K_2}+C_n $}

If the Grover coin is used at all vertices, the discrete time quantum walk over graph $ \overline{K_2}+C_n$ is periodic, with a period of 12. The walk has perfect state transfer after \(6+12m\) steps, where \(m \in
\mathbb{Z}_{\geq0}\), from one vertex of $\overline{K_2}$ to the other,
provided the initial state is an equal superposition of all coin states at the initial vertex.

The initial state is an eigenvector of the Grover coin, with eigenvalue 1, so is unchanged by the first coin operation. After the first step the state at each of the nodes of $C_n$ is of the form:

\vspace{-5mm}
\begin{center}
\begin{equation}
\frac{\gamma}{\sqrt{n|\gamma|}} \begin{pmatrix} 1\\0
  \\ 0\\0\end{pmatrix}  \text{where} \ \gamma \ \in \mathbb{C}
\end{equation}
\end{center}
\vspace{5mm}
After six steps of the walk perfect state transfer is achieved. From the symmetry
of $\overline{K_2} +C_n$ it can be seen that the evolution is periodic, with
period 12. The choice of coin operator is important in acheiving this
perfect state transfer, using only a DFT coin for quantum walks over
$\overline{K_2}+ C_n$ does not result in any notable transport
properties. For the walk over $\overline{K_2}+ C_n$ starting with an
equal superposition at one vertex of $\overline{K_2}$, the
probabilities on the nodes of $\overline{K_2}$ do not depend on $n$
when $n > 1$. The initial state is also important, as we now discuss. 

\begin{figure}[H]
\begin{center} 
\includegraphics[scale = 0.35]{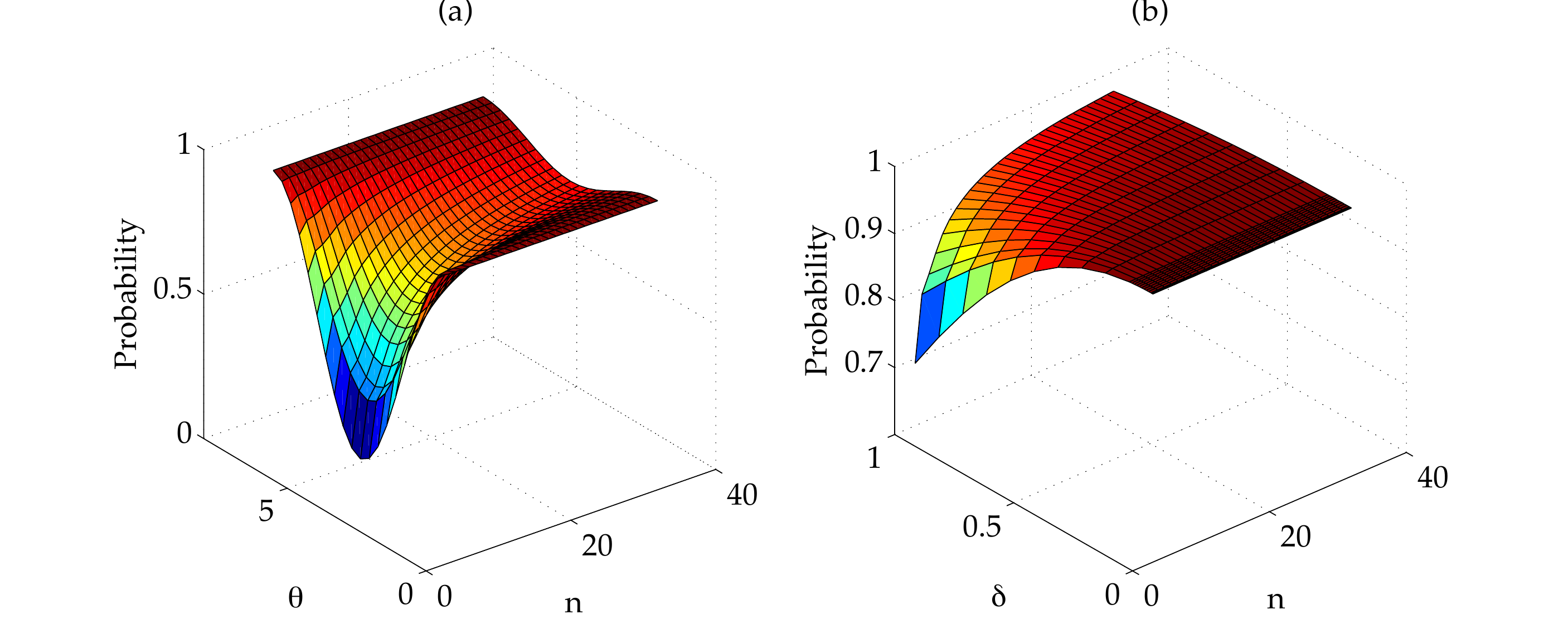}
\caption{\small Robustness of perfect state transfer to variation in
  a) initial phase and b) amplitude for one coin state. The size of
  $C_n$ is $n$,  $\theta $ is the phase and $\delta$ is the defect in
  Equation \ref{eqn:delta}. } \label{fig:robustness}
\end{center}
\end{figure}

\subsection{Robustness of perfect state transfer in graphs from $\overline{K_2}+C_n$ }

Numerical investigations were carried out to test the robustness of the first instance of perfect state transfer, after 6 steps, to variation in initial conditions. The transport was shown to be very robust to variations in initial condition of the form:\\
\begin{minipage}[c]{0.9\linewidth}
\begin{center} \begin{equation}  | \psi_{int} \rangle =
\frac{1}{\sqrt{(n- 2\delta+\delta^2)}}
\begin{pmatrix} 1\\1
  \\ \vdots \\1-\delta\end{pmatrix} ~\label{eqn:delta}
\end{equation} \end{center}
 \vspace{2mm} \end{minipage}\\
As $n$ increases, the robustness increases too, even in the case where $\delta = 1$, where the state is of a qualitatively different form to the standard initial state. Taking $\delta > 1$ adds a phase to the perturbed initial coin state. The robustness of the perfect state transfer to variation in phase of one of the initial coin states was also investigated, with similar findings, see Figure \ref{fig:robustness}. Though the perfect state transfer is much less robust to variation in phase, with probability going down to 0.19 for $n = 5$ when we have a phase factor of $\pi$, we can again increase the robustness of the transfer by increasing $n$. For $n > 35$ the probability at the target vertex after 6 steps does not go below 0.9 for any phase. The robustness increases with $n$ because an equal superposition amongst $n$ coin states is used for the initial state, and hence perturbing one part of that superposition results in a smaller relative perturbation as $n$ increases. If the initial state is instead varied by perturbing each coin state by a random $0\leq \delta \leq 1$ then the robustness decreases slightly with $n$. For $n = 3$ the amplitude at the target node goes down to 0.82, averaged from 1000 runs, at the perfect state transfer time. As $n$ increases the value tails off at 0.77.

\begin{center}
\begin{figure}[H]
\includegraphics[scale = 0.38]{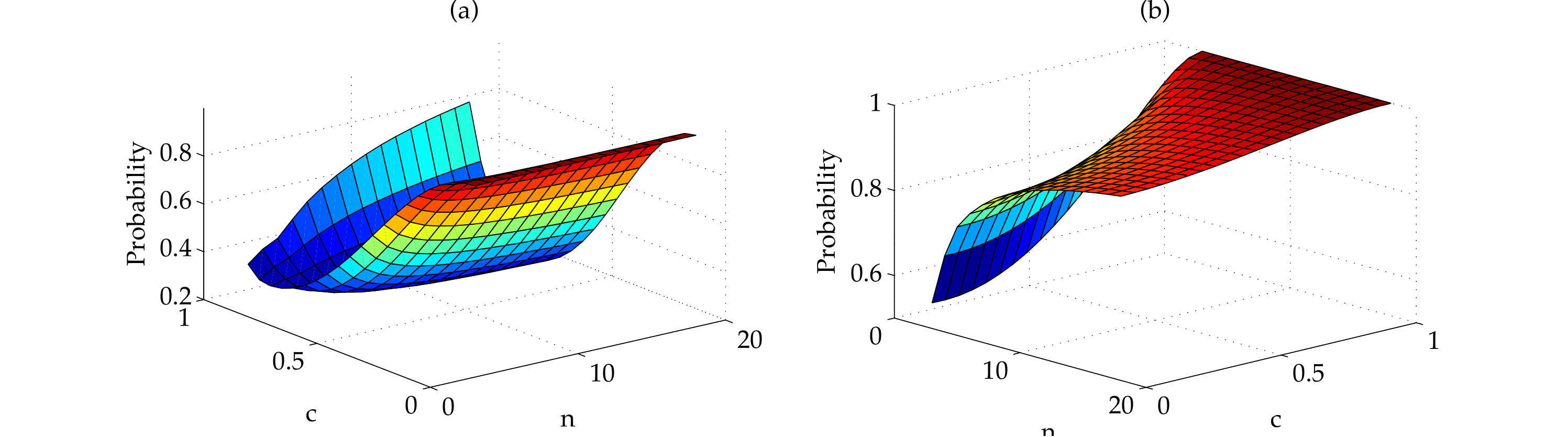}
\caption{Variation of fidelity of state transfer during Interpolation
  between a) Graphs $\overline{K_2}+\overline{K_n}$ and
  $\overline{K_2}+P_n$; and b)  $\overline{K_2}+P_n$ and
  $\overline{K_2}+ C_n$. $n$ is the number of nodes in $K_n$, $P_n$ or
$C_n$ and $c$ is the weighting of the edges that these graphs differ by. The orientations of the graphs are selected for clearest viewing.}
\label{fig:inter}
\end{figure}
\end{center}

The robustness of the transport with respect to decoherence was also tested. The effects of decoherence in the coin state and decoherence in the position state are both independent of $n$, and probability at the target vertex decays smoothly as the rate increases, recovering the classical distribution for decoherence rate $p = 1$ as expected.

\subsection{Interpolation between the three families}

It is possible to interpolate between graphs with the same vertices but
different edges by weighting the edges. To perform quantum walks on
such graphs, coins reflecting this edge weighting are required.
The coin should be the Grover operator of dimension $(d-t)$ when no edge is present, and that of dimension $d$ when the edge is fully present. Although walks using the Grover coin over both
$\overline{K_2}+\overline{K_n}$ and $\overline{K_2}+ C_n$ display
perfect state transfer, adding the edges they differ by gradually by
using a coin that interpolates between Grover operators of different
dimensions in the way described in \cite{Lovett11} destroys this
perfect state transfer. The coin is specified by

\begin{equation}
G_{d,t} = \begin{pmatrix} a \ & \ b \ & \ b \ & \ \ldots \ & \ b \ & \ c \ & \ c\ & \ c \ & \
    \ldots \ & \ c \\ b \ & \ a\ & \ b \ & \ \ldots \ & \ b \ & \ c \ & \ c \ & \ c \ & \ \ldots \ & \ c \\
    . \ & \ . \ & \ . \ & \ \ldots \ & \ . \ & \ . \ & \ . \ & \ . \ & \ \ldots \ & \ . \\
    . \ & \ . \ & \ . \ & \ \ldots \ & \ . \ & \ . \ & \ . \ & \ . \ & \ \ldots \ & \ . \\
    . \ & \ . \ & \ . \ & \ \ldots \ & \ . \ & \ . \ & \ . \ & \ . \ & \ \ldots \ & \ . \\ b \ & \ b \ & \ b \ & \
    \ldots \ & \ a \ & \ c \ & \ c \ & \ c \ & \ \ldots \ & \ c \\ c \ & \ c \ & \ c \ & \
    \ldots \ & \ c \ & \ e \ & \ f \ & \ f \ & \ \ldots \ & \ f \\  c \ & \ c \ & \ c \ & \
    \ldots \ & \ c \ & \ f \ & \ e \ & \ f \ & \ \ldots \ & \ f \\
    . \ & \ . \ & \ . \ & \ \ldots \ & \ . \ & \ . \ & \ . \ & \ . \ & \ \ldots \ & \ . \\
    . \ & \ . \ & \ . \ & \ \ldots \ & \ . \ & \ . \ & \ . \ & \ . \ & \ \ldots \ & \ . \\
    . \ & \ . \ & \ . \ & \ \ldots \ & \ . \ & \ . \ & \ . \ & \ . \ & \ \ldots \ & \ . \\ c \ & \ c \ & \ c \ & \
    \ldots \ & \ c \ & \ f \ & \ f \ & \ f \ & \ \ldots \ & \ e \end{pmatrix} 
\end{equation}

Where $a \ldots f \in \mathbb{R}$ where there are $(d-t)$ `normal' edges giving rise to blocks of size $(d-t)$ containing $a's$ and $b's$ and $t$ 'tunneling' edges with transitions specified by the block containing $e's$ and $f's$. This coin enables an edge to be `turned on' with strength $c$. The effect of turning on the additional edges to go from $\overline{K_2} + \overline{K_n}$ to $\overline{K_2}+ C_n$  does not depend on $n$. The amplitude at the target vertex at the perfect state transfer time decays quickly with $c$, and rises again very slowly as $c \rightarrow 1$. One can also interpolate between  $\overline{K_2}+\overline{K_n}$ and $\overline{K_2}+ C_n$ by going via $\overline{K_2}+P_n$. This interpolation does not improve the transport properties of the walk over $\overline{K_2}+P_n$, as can be seen in Figure \ref{fig:inter}.

\begin{center} 
\begin{figure}[H]
\centering \includegraphics[scale=0.47]{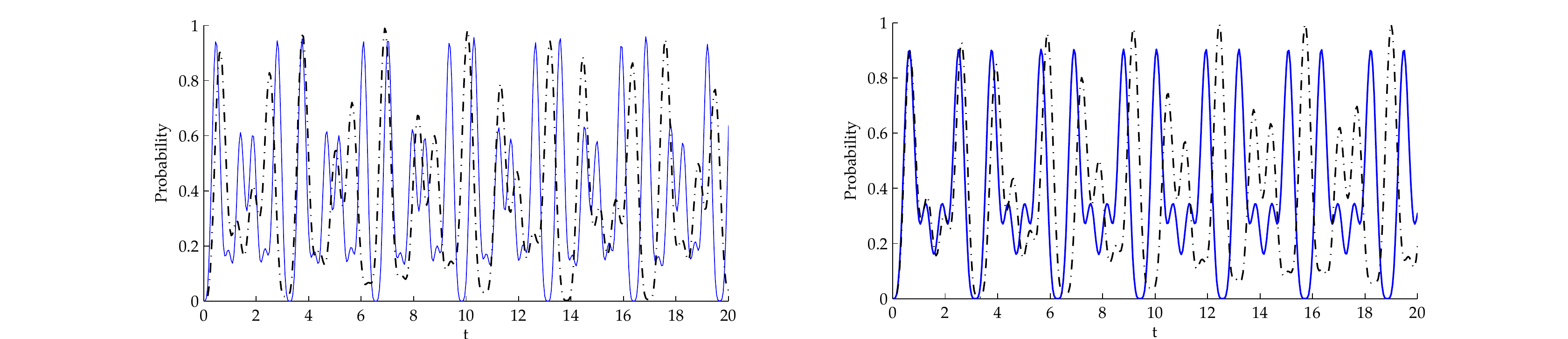}
\caption{Time evolution of probability at the target node for a) graph $\overline{K_2}+ P_n$ with $n = 11$ (solid) and $n=10$ (dotted); b) graph $\overline{K_2}+ P_n$ with $n=21$ (solid) $n = 11$ (dotted).}
\label{fig:fam3} 
\end{figure} 
\end{center}

\vspace{-10mm}
 \subsection{The continuous time walk on $\overline{K_2}+\overline{K_n}$, $\overline{K_2}+C_n$ and $\overline{K2}+P_n$}

In general, a graph exhibiting periodicity, high probability, or perfect state transfer for a specific coin in the discrete time walk is no indication that this will occur for the continuous time walk over the same graph. However for the three families of graphs discussed in this section, the continuous time walks exhibit perfect or very high fidelity state transfer. 

The continuous time walk on $\overline{K_2} + \overline{K_n}$ displays perfect state transfer for any $n$, and is hence periodic due to the symmetry of the graph. The perfect state transfer time can be tuned by adjusting $n$, as $n$ increases, the period decreases. An analytic expression for the period in terms of $n$ was deduced by inspection of the eigenvectors. The eigensystems of the graphs $\overline{K_2} + \overline{K_n}$ have only two nonzero eigenvalues of the form $\sqrt{2n}$, hence only two components of the analytic expression for the time evolution determine the period, regardless of the initial state used. The period of the graph is $2 \pi/\sqrt{2n}$, giving a perfect state transfer time between vertices of $\overline{K_n}$ of $\pi/\sqrt{2}$. The time evolution over $\overline{K_2}+ \overline{K_n}$ with initial state $| \psi \rangle = |n+1 \rangle$, with $n+1$ being one vertex of $\overline{K_2}$, is given by:\\
\vspace{-3mm}
\begin{center}
\begin{equation}
\label{eqn:timeev}
|\psi(t) \rangle = \left( \begin{array}{c} \frac{1}{\sqrt{2n}}(-i \ sin (\sqrt{2n} \ t)) \\
\vdots \\
\frac{1}{\sqrt{2n}}(-i \ sin (\sqrt{2n} \ t)) \\
\frac{1}{2}(cos(\sqrt{2n} \ t) + 1) \\
\frac{1}{2}(cos(\sqrt{2n} \ t) - 1) \end{array} \right) \end{equation}
\end{center}
\vspace{5mm}
where the first $n$ entries are the vertices of $\overline{K_n}$, the $(n+1)^{th}$ entry is the initial vertex, and the $(n+2)^{th}$ entry is the final vertex. This expression applies for any $n$, including $n= 0$ where the vertices of  $\overline{K_n}$ are not connected, so no evolution can occur. The cases for $n=1$ and $n=2$, giving rise to the graphs $P_3$ and $C_4$ respectively are already known \cite{Christandl04}.

There is little difference between the evolution over the graphs $\overline{K_2}+ P_n$ and $\overline{K_2}+ C_n$, as might be expected given that they differ by a single edge. Both exhibit oscillatory motion at the initial and target vertices, as depicted in Figure \ref{fig:fam3}, with the oscillations peaking at very high probabilities or unity. Clearly from the choice of $n$'s plotted, perfect state transfer does not occur for every $n$, so there is no simple relationship between the evolutions for different $n$'s. As for $\overline{K_2}+ \overline{K_n}$, the periods of the oscillations get smaller as $n$ increases. Decoherence in these walks quickly smears out the oscillatory behaviour, destroying the periodicity and perfect state transfer. 

\section{Discussion}
We investigated discrete time quantum walks over small graphs for a variety of choices of coin operation. The graphs were variations of graphs known to exhibit perfect state transfer under some circumstances, most notably diamond chains and cycles of length 4, 6 and 8. We have systematically investigated how a range of specific variations, namely the addition of up to four nodes, affects the perfect state transfer. We found that in general, varying the graphs in this way destroys the perfect state transfer. There are a small number of simple ways in which adding new nodes to the graph structure does not affect the perfect state transfer. These held for each of the cycles tested as they only involved modification of antipodal nodes. As long as the modifications do not add new paths between the antipodal nodes, the number of interim nodes in the cycle will only affect the number of timesteps required to obtain perfect state transfer for operator $\mathcal{O}_2$ rather than whether that transfer occurs. No dramatically new ways of achieving perfect state transfer were found, despite many of the structures investigated having properties shown analytically to be important for perfect state transfer- namely symmetry in the graph structures and coin operators \cite{Krovi,Krovi06}. As some of these results \cite{Krovi06} were obtained for graphs fulfilling very strict criteria, it is not surprising that these properties do not, in general lead to the specification of walks with perfect state transfer.

In terms of future work, if we intend to limit the size of the graphs examined, there is little point in adding further nodes to the graphs tested. The effects of adding extra connections between the existing vertices of the cycle were not examined. As all of the cases in which perfect state transfer was not destroyed relied on the fact that our operator $\mathcal{O}_2$ is simply a swap operator at nodes of degree 2, increasing the degree of these nodes does not appear to be a good way of varying the structure for the purposes of investigating perfect state transfer. A systematic study of all graphs upto a certain size may be required in order to obtain some potentially more interesting results. Whilst there is also the freedom to vary the coin operators, our choices are natural as they arise in many discussions of the quantum walks \cite{Tregenna03,Groverwalk,Kempe05,Kendon07}. 

Though many of the graphs exhibited high amplitude transfer to some node for a limited subset of initial conditions tested, we were looking for perfect state transfer between antipodal nodes occurring for a variety of initial conditions. Some of the graphs found to have perfect state transfer between antipodal nodes could be generalised into families of graphs, and the dynamics of quantum walks on these families were investigated in detail. Two of the families of graphs, for the right choices of coin operator in the discrete time walk, exhibit periodicity, and in some cases perfect state transfer. We found that these families also exhibit very high amplitude transfer, or perfect state transfer, in the continuous time case. In the case of the discrete time walk over $\overline{K_2}+ C_n$ it was found that increasing $n$ improved robustness of the perfect state transfer to variations in one coin state of the initial state, but not to decoherence. In the continuous time case, $n$ can be used to tune the perfect state transfer time. 

Characterising the graphs which admit perfect state transfer remains an ongoing project, as is understanding which structural properties of graphs give rise to perfect state transfer. We have found cases where adding an arbitrary number of nodes to a particular graph in the right way preserves its state transfer properties. Ideally general methods of determining whether a graph has perfect state transfer, such as that outlined in \cite{Godisil11} for the continuous time case, are required. Due to the coin degrees of freedom, and the choice in coin operator, this is a far more difficult task in the discrete time case. 


{ \small KB and DA were funded by the UK Engineering and Physical
Sciences Research Council, VK is funded by the a UK Royal
Society University Research Fellowship, and TP was funded by
a UK Royal Society Summer project bursary.}
 
\newpage
\appendix
\hspace{-6mm}\textbf{Appendix}
\section{}
\label{sec:appa}
Summary of the numerical results relating to state transfer to the end of a diamond chain, results for $4 \leq n < 10 $ are omitted as they follow the trend indicated in Section \ref{sec:diamond}:\\
\vspace{1mm}

\begin{tabular}{| c | c | c | c | c | c | c |}
\hline
Coin & $n$ & Chain (a) max & Chain (b) max & Chain (c) max & Chain (d) max \\ \noalign{\hrule height 1.2pt}
$\mathcal{O}_1$ & 2 & 0.52 & 0.52 & 0.24 & 0.30 \\ \hline
     & 3 & 0.27 & 0.23 & 0.18 & 0.33 \\ \hline
     & 10 & 0.05 & 0.04 & 0.06 & 0.26 \\ \noalign{\hrule height 1.2pt}
$\mathcal{O}_2$ & 2 & 1.00 & 0.99 & 0.69 & 0.77 \\ \hline
     & 3 & 1.00 & 0.99 & 0.70 & 0.46 \\ \hline
     & 10 & 1.00 & 0.98 & 0.12 & 0.23 \\ \noalign{\hrule height 1.2pt}
$\mathcal{O}_3$ & 2 & 0.35 &  0.25 & 0.78 & 0.71 \\ \hline
     & 3 & 0.14 & 0.14 & 0.46 & 0.64 \\ \hline
     & 10 &  0.07 & 0.13 & 0.22 & 0.20 \\ \noalign{\hrule height 1.2pt}
\end{tabular}\\
The perfect state/high amplitude transfer for chains (a) and (b) using operator $\mathcal{O}_2$ occur after $2n$ timesteps. 
\vspace{3mm}

\section{}
\label{sec:appb}

\begin{figure}  
\subfigure a) {\includegraphics[scale = 0.75]{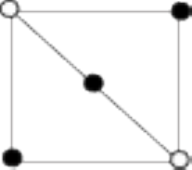}} 
\subfigure b) {\includegraphics[scale = 0.75]{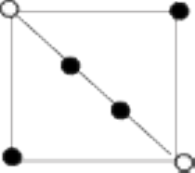}} 
\subfigure c) {\includegraphics[scale = 0.75]{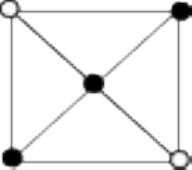}} 
\subfigure d) {\includegraphics[scale = 0.75]{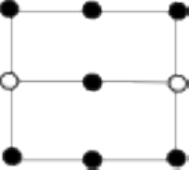}} 
\subfigure e) {\includegraphics[scale = 0.75]{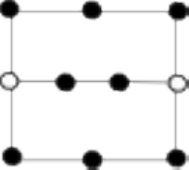}} \\ 
\subfigure f) {\includegraphics[scale = 0.75]{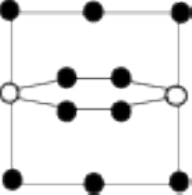}} 
\subfigure g) {\includegraphics[scale = 0.75]{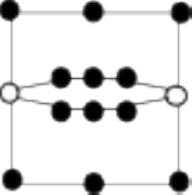}} 
\subfigure h) {\includegraphics[scale = 0.75]{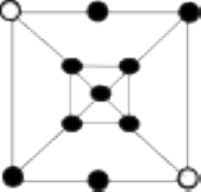}} 
\subfigure i) {\includegraphics[scale = 0.75]{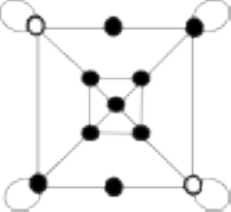}} 
\subfigure j) {\includegraphics[scale = 0.75]{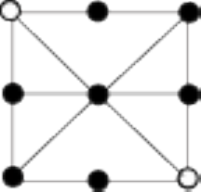}} 
\subfigure k) {\includegraphics[scale = 0.76]{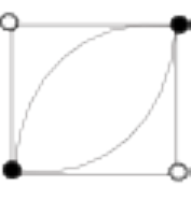}} 
\caption{Variants of 4 cycles: a), b), c), d), k), 6 cycles: (h), (i) and 8 cycles: e), f), g), j) preliminarily investigated. Open circles indicate the starting and end nodes, and due to the symmetries of the graphs these are interchangeable.} \label{fig:cycles} \end{figure} 

Summary of the maximum state transfer achieved numerically between highlighted nodes of graphs in Figure \ref{fig:cycles}:\\
\vspace{1mm}

\begin{tabular}{| c | c | c | c |}
\hline
  Graph &   Coin (1) max   &  Coin (2) max  & Coin (3) max  \\ \hline
  (a) & 0.99 & 0.96 & 0.96 \\ \hline
  (b) & 0.97 & 0.96 & 0.96 \\ \hline
  (c) & 0.97 & 0.96 & 0.96\\ \hline
  (d) & 0.83 & 1.00 & 0.90 \\ \hline
  (e) & 0.42 & 1.00 & 0.70 \\ \hline
  (f) & 0.62 & 0.95 & 0.48\\ \hline
  (g) & 0.51 & 0.98 & 0.60\\ \hline
  (h) & 0.44 & 0.75 & 0.75\\ \hline
  (i) & 0.44 & 0.77 & 0.77 \\ \hline
  (j) & 0.20 & 0.40 & 0.40 \\  \hline 
  (k) & 0.74 & 0.73 & 1.00 \\ \hline
\end{tabular} \\
Typically the high amplitude transfer observed is very sensitive to initial conditions, for all variants but (a) with high amplitude transfer fewer than 3\% of initial states tested exhibited this property.

\newpage
\bibliographystyle{unsrt}
\bibliography{bibitems2}

\end{document}